\documentclass[10pt]{article}

\usepackage{amsfonts}
\usepackage[centertags]{amsmath}
\usepackage{amssymb}
\usepackage{cite}
\usepackage{psfrag}
\usepackage{graphicx}
\usepackage{bbm,bm}
\usepackage{epsfig, multicol}
\usepackage{setspace}
\allowdisplaybreaks
\begin{document}

\topmargin 0pt
\oddsidemargin 0mm
\def\be{\begin{equation}}
\def\ee{\end{equation}}
\def\bea{\begin{eqnarray}}
\def\eea{\end{eqnarray}}
\def\ba{\begin{array}}
\def\ea{\end{array}}
\def\ben{\begin{enumerate}}
\def\een{\end{enumerate}}
\def\nab{\bigtriangledown}
\def\tpi{\tilde\Phi}
\def\nnu{\nonumber}
\newcommand{\eqn}[1]{(\ref{#1})}

\newcommand{\half}{{\frac{1}{2}}}
\newcommand{\vs}[1]{\vspace{#1 mm}}
\newcommand{\dsl}{\pa \kern-0.5em /} 
\def\a{\alpha}
\def\b{\beta}
\def\g{\gamma}\def\G{\Gamma}
\def\d{\delta}\def\D{\Delta}
\def\ep{\epsilon}
\def\et{\eta}
\def\z{\zeta}
\def\t{\theta}\def\T{\Theta}
\def\l{\lambda}\def\L{\Lambda}
\def\m{\mu}
\def\f{\phi}\def\F{\Phi}
\def\n{\nu}
\def\p{\psi}\def\P{\Psi}
\def\r{\rho}
\def\s{\sigma}\def\S{\Sigma}
\def\ta{\tau}
\def\x{\chi}
\def\o{\omega}\def\O{\Omega}
\def\k{\kappa}
\def\pa {\partial}
\def\ov{\over}
\def\nn{\nonumber\\}
\def\ud{\underline}
\def\qq{$Q{\bar Q}$}
\begin{flushright}
%
\end{flushright}
\begin{center}
{\Large{\bf Holographic entanglement entropy, subregion complexity and Fisher information metric of `black' non-susy D3 Brane}}

\vs{10}

{Aranya Bhattacharya\footnote{E-mail: aranya.bhattacharya@saha.ac.in} 
and Shibaji Roy\footnote{E-mail: shibaji.roy@saha.ac.in}}

\vs{4}

{\it Saha Institute of Nuclear Physics\\
1/AF Bidhannagar, Calcutta 700064, India}

\vs{4}

{\rm and}

\vs{4}

{\it Homi Bhabha National Institute\\
Training School Complex, Anushakti Nagar, Mumbai 400085, India}
\end{center}

\vs{10}

\begin{abstract}
The BPS D3 brane has a non-supersymmetric cousin, called the non-susy D3 brane, which is also a solution of type
IIB string theory. The corresponding counterpart of black D3 brane is the `black' non-susy D3 brane and like the
BPS D3 brane, it also has a decoupling limit, where the decoupled geometry (in the case we are interested, this is 
asymptotically AdS$_5$ $\times$ S$^5$) is the holographic dual of a non-conformal,
non-supersymmetric QFT in $(3+1)$-dimensions. In this QFT we compute the entanglement entropy (EE), the complexity and the 
Fisher information metric holographically using the above mentioned geometry for spherical 
subsystems. The fidelity and the Fisher information metric have been calculated from the regularized extremal 
volume of the codimension one time slice of the bulk geometry using two different proposals in the literature. Although for AdS black hole both the proposals give identical results, the results differ for the non-supersymmetric background.
\end{abstract}

\newpage

\section{Introduction}

In recent years, significant amount of work has been done to understand the gravity duals of certain measures of quantum 
information \cite{ Witten:2018zva}, 
namely, the entanglement entropy (EE) \cite{Bombelli:1986rw, Srednicki:1993im, Holzhey:1994we, Calabrese:2004eu, 
Calabrese:2005zw, Calabrese:2009qy, 
Nishioka:2009un, Takayanagi:2012kg, Witten:2018lha}, the fidelity susceptibility or Fisher information metric 
\cite{UHLMANN1976273, 
Hayashi, Petz, Lashkari:2015hha, MIyaji:2015mia, Alishahiha:2017cuk, Banerjee:2017qti}, the Bures metric
\cite{PhysRevLett.72.3439} and so on.  
The AdS/CFT correspondence \cite{Maldacena:1997re, Aharony:1999ti} appears to be the most useful tool for this purpose. 
The primary motivation in this came from the 
seminal work by Ryu and Takayanagi \cite{Ryu:2006bv, Ryu:2006ef}, where they gave a proposal to quantify the EE 
holographically, particularly, 
in case of spacetimes with constant negative curvature. Results obtained using their holographic proposal matched 
exactly with those of the corresponding CFT duals in low dimensions. 
As it is quite well-known by now, the EE is a good way to measure the amount of quantum information 
of a bipartite system. One way to quantify this information is to calculate the von Neumann entropy of a bipartite system 
where the system is divided into two parts named, $A$ and $B$. The von Neumann entropy of part $A$ is defined as $S_A = - {\rm Tr}(\rho_A \log \rho_A)$,
where $\rho_A = {\rm Tr}_B (\rho_{\rm tot})$ is the reduced density matrix on $A$ obtained by tracing out on $B$, the complement of $A$,
of the density matrix of the total system $\rho_{\rm tot}$. Holographically it can be computed from 
the Ryu-Takayanagi formula (as proposed by them) \cite{Ryu:2006bv, Ryu:2006ef}
\be\label{HEE}
S_E = \frac{{\rm Area}(\gamma_A^{\rm min})}{4G_N}
\ee
where $\gamma_A^{\rm min}$ is the $d$-dimensional minimal area (time-sliced) surface in AdS$_{d+2}$ space whose boundary matches with 
the boundary of the subsystem $A$,
i.e., $\partial\gamma_A^{\rm min} = \partial A$ and $G_N$ is the $(d+2)$-dimensional Newton's constant. As mentioned earlier, for lower 
spacetime dimensions (AdS$_{3}$/CFT$_{2}$) the corresponding results matched. Since then, this dual description has been 
checked for several cases and it's regime of application has been extended to cases of higher dimensional and asymptotically 
AdS spacetimes \cite{Bhattacharya:2012mi, Allahbakhshi:2013rda, Bhattacharya:2017gzt}. 
For asymptotic AdS cases, one finds extra finite contributions to the EE other than that of the pure AdS spacetimes 
which has also been studied in details in several works. These terms are found to follow certain relations analogous
to the thermodynamical relations called, the entanglement thermodynamics \cite{Bhattacharya:2012mi, Allahbakhshi:2013rda,Pang:2013lpa, Chakraborty:2014lfa}.

Complexity is another measure of entanglement between quantum systems and a holographic definition in the context of 
eternal AdS black hole \cite{Maldacena:2001kr}
was originally proposed by Susskind et. al. \cite{Susskind:2014rva,  Stanford:2014jda ,
 Brown:2015bva, Brown:2015lvg} through two different notions, one from the 
volume of the Einstein-Rosen bridge (ERB) connecting the two boundaries
of the black hole and the other is by the action of its Wheeler-DeWitt patch as given below,
\begin{equation}
C_{V} = \left(\frac{V(\gamma)}{R G_{N}}\right), \qquad \qquad C_{A}=\frac{I_{WDW}}{\pi \hbar}
\end{equation}
where $R$ is the AdS radius, $V(\gamma)$ is the maximum volume of the co-dimension one bulk surface (time-sliced) bounded by the two
boundaries of the black hole and $I_{WDW}$ is the action of the Wheeler-DeWitt patch.

Motivated by Susskind et. al., another definition of holographic complexity has been proposed by Alishahiha \cite{Alishahiha:2015rta} by the
codimension one volume of the time-slice of the bulk geometry enclosed by the extremal codimension two Ryu-Takayanagi (RT) hypersurface
used for the computation of holographic EE. This is usually referred to as the subregion complexity 
\cite{Ben-Ami:2016qex, Carmi:2016wjl, Roy:2017kha} and the relation
between these two notions has been clarified in some recent works given in 
\cite{Agon:2018zso,Alishahiha:2018lfv,Takayanagi:2018pml}. The subregion complexity, which we calculate 
in this paper, is defined in a very similar way as, $C_{V} = \frac{V_{RT}(\gamma)}{8\pi R G_N}$, where $V_{RT}$ denotes the
volume enclosed by the RT surface. This is closely related to quantum complexity, a concept borrowed from quantum
information theory. Formally, complexity of a CFT state (corresponding to
a particular quantum circuit) is defined as the minimum number of simple unitaries (or gates) required to prepare the state 
from a fixed reference state. So, if $|\psi(R)\rangle$ denotes the reference state and $|\psi(T)\rangle$ denotes the final
target state, then complexity of $|\psi(T)\rangle$ with respect to the reference state would be the minimum number of
quantum gates to form a single unitary $U(R,T)$ satisfying the equation,
\be
|\psi(T)\rangle = U(R,T) |\psi(R)\rangle
\ee
with some tolerance or limit of how close one can reach to the target state.     
However, a clear field theory description of holographic complexity is not yet known (see, however, \cite{Caputa:2017urj,
Caputa:2017yrh, Abt:2017pmf}). 
Using the geometric approach of Nielsen to the quantum circuit model \cite{Nielsen1, Nielsen2}, the complexity of only some free field theory states have 
been found to resemble the holographic complexity in \cite{Chapman:2017rqy,Jefferson:2017sdb,Khan:2018rzm,Hackl:2018ptj}. But, for interacting field theory this 
is far from clear. Many different descriptions and proposals have been 
given in the literature to relate holographic complexity to the fidelity susceptibility or the quantum Fisher information metric, the Bures metric 
and so on \cite{MIyaji:2015mia, Alishahiha:2017cuk, Banerjee:2017qti}. In the same 
way as EE, complexity can be calculated perturbatively in case of asymptotically AdS solutions. One of the recent proposals
relates the holographic complexity as the dual to quantum Fisher 
information metric \cite{Banerjee:2017qti}. It relates the regularized
second order change of RT volume with respect to a perturbation parameter to the fidelity $\mathcal{F} \sim (V_{RT}^{m^{2}}-V_{AdS})$, 
where $m$ is a perturbation parameter. The corresponding Fisher information metric has been proposed to have a form 
\begin{equation}
G_{F,mm}= \partial^{2}_m \mathcal{F}
\end{equation}

In this paper, we work with the decoupled geometry of `black' non-susy D3 brane solution of type IIB string theory
\cite{Lu:2007bu, Nayek:2016hsi, 
Chakraborty:2017wdh, Bhattacharya:2017gzt}. This geometry is the gravity
dual of a non-supersymmetric, non-conformal QFT in (3+1) dimensions at finite temperature which is also confining and has running coupling
constant very much like QCD. As we will see the geometry is asymptotically AdS$_5$ which means that it can be thought of as some 
non-supersymmetric and non-conformal deformation of the CFT which is ${\cal N}=4$, $D=4$ SU($N$) super Yang-Mills theory at large $N$.
We compute the EE, complexity and Fisher information metric holographically in this background for
spherical subsystems. The goal of this study would be to gain better understanding of the various phases of QCD-like theories and the 
transitions among them since it is believed that the EE and the complexity are possibly related to some universal properties like 
order paramater or some renormalization group flow \cite{Chapman:2017rqy}. However, this will be more clear once we have a clearer 
picture of the holographic complexity in the (strongly coupled) interacting field theories. 

The non-susy D3-brane solution is given by the following
field configurations consisting of a Einstein-frame metric, a dilaton and a self-dual 5-form field strength,   
\bea\label{nonsusyd3}
& &  ds^{2}= F_1(\rho)^{-\frac{1}{2}}G(\rho)^{-\frac{\delta_{2}}{8}}\left[-G(\rho)^{\frac{\delta_{2}}{2}}dt^{2}+ \sum_{i=1}^{3}(dx^{i})^{2}\right]+F_1(\rho)^{\frac{1}{2}}
G(\rho)^{\frac{1}{4}}\left[\frac{d\rho^{2}}{G(\rho)}+\rho^{2}d\Omega_{5}^{2}\right]\nn
& & e^{2\phi} = G(\rho)^{-\frac{3\d_2}{2} + \frac{7\d_1}{4}}, \qquad
 \qquad F_{[5]} = \frac{1}{\sqrt{2}}(1+\ast) Q {\rm Vol}(\Omega_5).
\eea
where the functions $G(\rho)$ and $F(\rho)$ are defined as,
\be\label{functions}
G(\rho)=1+\frac{\rho_{0}^{4}}{\rho^{4}},\qquad
F_1(\rho)=G(\rho)^{\frac{\alpha_1}{2}}\cosh^{2}\theta - G(\rho)^{-\frac{\beta_1}{2}}\sinh^{2}\theta 
\ee
Here $\d_1$, $\d_2$, $\a_1$, $\b_1$, $\theta$, $\rho_0$, $Q$ are the parameters characterizing the solution. The parameters satisfy
$\a_1=\b_1=(1/2) \sqrt{10 - (21/2)\d_2^2 - (49/4) \d_1^2 + 21 \d_1\d_2}$ and $Q = 2\a_1\rho_0^4 \sinh 2\theta$. For simplicity,
we put $\a_1+\b_1=2$ (in this case S$^5$ has a constant radius), implying $\a_1=\b_1=1$ and $\d_1$, $\d_2$ satisfy $42\d_2^2+49\d_1^2-84\d_1\d_2=24$. 
The decoupled geometry can be obtained by zooming into the region $\rho \sim \rho_0 \ll \rho_0\cosh^{\half}\theta$ and the metric in \eqref{nonsusyd3} 
then reduces to the form (without the product S$^5$ part),      
\be\label{aads5}
ds^{2}= \frac{\rho^2}{R_1^2}G(\rho)^{\frac{1}{4}-\frac{\delta_{2}}{8}}\left[-G(\rho)^{\frac{\delta_{2}}{2}}dt^{2}+ \sum_{i=1}^{3}(dx^{i})^{2}\right]+
\frac{R_1^2}{\rho^2}\frac{d\rho^{2}}{G(\rho)}
\ee
This geometry is asymptotically ($\rho \to \infty$) AdS$_5$ with radius $R_1 = \rho_0\cosh^{\half}\theta$ and so, it can be regarded as
an excited state over AdS$_5$. The solution reduces to AdS$_5$ black hole once we take $\d_{2}=-2$ and $\d_1 = -12/7$. In general,
the non-susy solution can be shown \cite{Kim:2007qk} to have a temperature 
$T_{\rm nonsusy} = \left(-\frac{\d_2}{2}\right)^{1/4} \frac{1}{\pi \rho_0 \cosh\theta}$ (although
there is a singularity at $\rho = 0$) which gives standard AdS$_5$ black hole temperature for $\d_2=-2$. In our earlier work \cite{Bhattacharya:2017gzt}
we computed the 
change in EE for this asymptotically AdS$_5$ state and obtained the entanglement thermodynamics upto first order (in perturbation parameter $m$),
considering only
a small strip type subsystem. In this paper we compute holographically the change in EE as well as the change in complexity 
in both the first and the second order for the small spherical subsystem. Here $m \equiv \frac{1}{z_0^4} \equiv \frac{\rho_0^4}{R_1^8}$
will be treated as the perturbation parameter we mentioned before. Also as noted earlier, from the second order change in complexity which is
related to the second order change in regularized RT volume, we obtain the form of fidelity and Fisher information metric of the boundary QFT. 

The rest of the paper is organized as follows. 
The holographic EE and the complexity computation of AdS$_5$ black hole
for spherical subsystem have been reviewed in section 2. 
In section 3, we give the holographic EE and the holographic complexity 
for decoupled `black' nonsusy D3 brane for spherical subsystem.  Finally we conclude in section 4.

\section{EE and complexity of AdS$_5$ black hole for spherical subsystem}

The AdS$_5$ black hole geometry can be obtained from the decoupling limit of non-extremal D3-brane solution of type IIB
string theory \cite{Witten:1998zw} and the metric has the form
\begin{equation}
ds^{2}=\frac{R_1^{2}}{z^{2}} \left[-\left(1-\frac{z^{4}}{z_{0}^{4}} \right)dt^{2}+ \sum_{i=1}^3 (dx^i)^2 + \frac{dz^{2}}{\left(1-\frac{z^{4}}{z_{0}^{4}}\right)}\right]
\end{equation}
where $z=z_0$ is the location of the horizon. Note that this metric reduces to AdS$_5$ metric for $z \to 0$ and therefore it is asymptotically AdS.
Now replacing $\frac{1}{z_{0}^{4}}$ by $m$, the metric takes the form 
\begin{equation}\label{adsbh}
ds^{2} = \frac{R_{1}^{2}}{z^{2}}\left[-\left(1-m z^{4} \right)dt^{2}+ \sum_{i=1}^{3}(dx_{i})^{2} + \frac{dz^{2}}{\left(1-m z^{4} \right)}\right]
\end{equation}
In the following we give the metric by rewriting the three dimensional Euclidean part in spherical
polar coordinates as,
\be\label{adsbh1}
ds^{2}= \frac{R_1^2}{z^2}\left[-\left(1-m z^{4} \right)dt^{2}+ \frac{dz^{2}}{\left(1-m z^{4} \right)} + dr^{2}+ r^{2}d\Omega_{2}^{2}\right]
\ee
This is very similar to the metric studied earlier in \cite{Bhattacharya:2012mi} and \cite{Alishahiha:2015rta}. Here, we briefly review 
their calculation of holographic EE 
and complexity\footnote{Here we note that in \cite{Alishahiha:2015rta} some generalities for the change in complexity of AdS black hole
for the spherical subsystem is given. Only for $d=1$ and $d=2$, the explicit results have been given. But since here we consider $d=3$, we
derive the change in complexity for spherical subsystem later in this section.} 
for the purpose of the calculation of the same
in more general `black' non-susy D3 brane background to be discussed in section 3.  We also compute the second order change 
of both  holographic EE and holographic subregion complexity.
The subsystem in this case is given by a round ball ($r^2 = \sum_{i=1}^{3}x_{i}^{2}\leq \ell^{2}$) on the boundary. The embedding of the surface in the bulk 
is specified by $r=r(z)$. The area of the embedded surface can be written as (we assume, $m \ell^{4}\ll 1$, where $m$ is related to the black hole horizon $z_{h}$
by the relation $m z_{h}^{4}=1$), 
\begin{equation}\label{area1}
A_{BH} = 4 \pi R_{1}^{3} \int_{z=\epsilon}^{L}\frac{dz}{z^{3}}r(z)^{2}\sqrt{\frac{1}{1-m z^{4}}+ r^{\prime}(z)^{2}}
\end{equation}
Here $L$ is a parameter (the turning point of the RT surface) which is closely related to the radius of the sphere $\ell$
(the turning point in case of pure AdS$_{5}$).
The entanglement entropy would be calculated from the area \eqref{area1} upto second order 
in $m$. We try to find the functional form of $r(z)$ by solving the Euler-Lagrange equation after expanding
it upto second order in $m$. For this we work with the ansatz
\be\label{ansatz}
r(z) = \sqrt{L^2 - z^2} + m r_{1}(z) + m^{2} r_{2}(z)
\ee
and solve the differential equation (obtained from the Euler-Lagrange equation for $r(z)$) of $r_{1}(z)$ and $r_{2}(z)$
perturbatively. The boundary condition we use is: 
Lim$_{z \to L} r_{1,2}(z) = 0$. We, therefore, first solve $r_{1}(z)$ which has the following form,
\begin{equation}\label{Rz}
    r_{1}(z)= \frac{2 L^6-z^4 \left(z^2+L^2\right)}{10 \sqrt{\left(L^2-z^2\right)}}
\end{equation}
Now using this, we can solve for $r_{2}(z)$ which is of the form
\begin{equation}
 r_{2}(z)= \frac{\sqrt{L^2-z^2} \left(464 L^8+380 L^6 z^2+228 L^4 z^4+328 L^2 z^6+175 z^8\right)}{4200}
\end{equation}

Using this form, we first get a relationship between $\ell$ and $L$,
by taking $r(z=0)=\ell$ as,
\begin{equation}\label{lL}
\ell=r(0)=L+ \frac{L^5 m}{5}+ \frac{58 L^9 m^2}{525},
\end{equation}
and the inverse relation thus looks like
\begin{equation}\label{inverse}
 L=  \ell - m \ell^5/5+\frac{47 \ell^9 m^2}{525}.
\end{equation}

Then using \eqref{ansatz} with derived versions of $r_1 (z)$ and $r_2 (z)$ in the area integral \eqref{area1} and expanding it upto the second order in $m$ we get the minimal area of RT
surface as
\bea
& & A_{0(BH)}=4\pi R_{1}^{3}\int_{\epsilon}^{L}dz \frac{L \sqrt{L^2-z^2}}{z^3}\\
& & A_{1(BH)}=4\pi R_{1}^{3} m \int_{\epsilon}^{L}dz \, \frac{  L \sqrt{L^2-z^2} \left(4 L^4+2 L^2 z^2+9 z^4\right)}{10 z^3}\\
& & A_{2(BH)} = 4\pi R_{1}^{3} m^2\int_{\epsilon}^{L}dz \, \frac{L \sqrt{L^2-z^2}}{4200z^3} \left[1096 L^8+548 L^6 z^2+1608 L^4 z^4 +1256 L^2 z^6+3367 z^8\right]
\eea 
Note that here $A_{0,1,2}$ are not the true zero, first and second order forms of the area due to the appearance of $L$
which also has an expansion in $m$ due to the inverse relation given in \eqref{inverse}. So, to get the correct forms of area at different orders we evaluate the sum of the
three integrals $A_{0(BH)}+A_{1(BH)}+A_{2(BH)}$ and then use the expansion of $L$. Finally, we put $\epsilon \to 0$. This way the zeroth order term will be the pure AdS result which is divergent, but the first and second order terms are finite and give the first and second order change in area.
Using these the first and second order change in holographic entropy can be written in the form
\begin{equation}\label{entropy1}
\Delta S_{EE(BH)}^{(1)}=\frac{\pi R_{1}^{3}}{10 G_{5} }m \ell^{4} 
\end{equation}
and
\begin{equation}\label{entropy2}
\Delta S_{EE(BH)}^{(2)}=-\frac{\pi R_{1}^{3}}{525 G_{5} }m^{2} \ell^{8}.
\end{equation}
We note that \eqref{entropy1} is precisely the relation obtained in \cite{Bhattacharya:2012mi} for the first order change in EE,
on the other hand, \eqref{entropy2} matches with the second order results obtained in \cite{Blanco:2013joa}.

Now we will extend this calculation to compute the holographic complexity for AdS$_5$ black hole. The volume integral here takes the form
\begin{equation}
V_{BH}=\frac{4\pi R_{1}^{4}}{3}\int_{\epsilon}^{L} dz \frac{r(z)^{3}}{z^4 \sqrt{1-m z^{4}}} 
\end{equation}
As before we replace $r(z)$ by the functional form given in \eqref{ansatz} after using $r_1$ and $r_2$. Then we expand the integral upto 
the second order\footnote{Note that for the computation of the second order change in holographic EE, it is actually enough to work with 
only the first order change in embedding, i.e., to take $r(z)$ upto 
$r_{1}(z)$, but this is not the case for the computation of the second order change in subregion complexity. Here we necessarily have to take 
the second order change in embedding, i.e., we have to take $r(z)$ upto $r_2(z)$.} in $m$. The zeorth, the first and the second order terms of 
the integral are given respectively as, 
\bea
& & V_{0(BH)}=\frac{4\pi R_{1}^{4}}{3}\int_{\epsilon}^{L}dz \frac{(L^2-z^2)^{3/2}}{z^4} =V_{AdS}\\
& & V_{1(BH)}= \left(\frac{4 \pi R_{1}^{4}}{15 }\right)m\int_{\epsilon}^{L}\frac{dz}{z^4}\sqrt{L^2-z^2} \left(3 L^6+L^2 z^4-4 z^6\right)\\
& & V_{2(BH)}= \left(\frac{4 \pi R_{1}^{4}}{1050}\right)m^2\int_{\epsilon}^{L}\frac{dz}{z^4}\left(L^2-z^2\right)^{1/2} 
\left(158 L^{10}+21 L^8 z^2+67 L^6 z^4\right.\nn
& & \qquad\qquad\qquad\qquad\qquad\qquad\qquad \left.-17 L^4 z^6+9 L^2 z^8-238 z^{10} \right)
\eea 
Here again, the 
integrals are taken care of in the way mentioned previously in case of area integrals. In case of 
non-supersymmetric solution,
we would not repeat this statement anymore. But the calculations are done in the same way. After all these steps one find that the first order change in volume
in the $\epsilon \rightarrow 0$ limit is zero. The second order change in complexity/volume can be
obtained from the second order change of regularized Ryu-Takayanagi volume which we write as, 
\be\label{changevol}
\Delta V_{(BH)}^{(2)}= \frac{4 \pi R_{1}^{4}}{3}\left(\frac{3\pi}{1280}\right) (m \ell^{4})^{2}
 \ee
and so the second order change in complexity for AdS$_5$ black hole is given by, 
\be\label{complexbh}
\Delta C_{V(BH)}^{(2)} = \frac{4\pi R_{1}^{3}}{24 \pi G_{5}}\left(\frac{3\pi}{1280}\right) (m \ell^{4})^{2} = \frac{\pi R_1^3}{2560 G_5}(m\ell^4)^2
\ee
Similar results have been obtained in \cite{Alishahiha:2015rta} only for AdS$_3$ and AdS$_4$ black hole with spherical subsystem.
Now for spherical subsystem the change of energy is given as \cite{Balasubramanian:1999re, deHaro:2000vlm, Allahbakhshi:2013rda},
\begin{equation}\label{energy}
\Delta E = \frac{4 \pi \ell^{3}}{3}\langle T_{tt}\rangle
  \end{equation}
Using $\langle T_{tt}\rangle =\frac{3R_{1}^{3}m}{16 \pi G_{5}}$, in \eqref{energy} we get,
\begin{equation}\label{deltae}
      \Delta E = \frac{R_{1}^{3}m\ell^{3}}{4 G_{5}}
  \end{equation}
Thus the entanglement temperature computed from \eqref{entropy2} and \eqref{deltae} comes out to be 
\begin{equation}
    T_{ent(BH)}= \frac{5}{2\pi \ell}
\end{equation}
consistent with the results of \cite{Bhattacharya:2012mi}. The change in entanglement entropy can be written as 
$\Delta S_{EE(BH)}^{(1)} = \Delta E / T_{ent(BH)}$, similarly,
one can write $\Delta C_{V(BH)}^{(2)} \sim (\Delta E / T_{ent(BH)})^2$ as, 
\begin{equation}
    \Delta C_{V(BH)}^{(2)} =\frac{5 G_5}{128 \pi R_1^3} \left(\frac{\Delta E}{T_{ent(BH)}}\right)^2
\end{equation}

Once we have the change in Ryu-Takayanagi volume (regularized) for an excited state, we can easily compute the fidelity and from
there obtain the Fisher information metric following the proposal of \cite{Banerjee:2017qti}. In general $d+1$ dimensions the change 
in Ryu-Takayanagi volume and the corresponding fidelity are given as,
\bea
& & \Delta V^{(2)} = \frac{R_1^d \Omega_{d-2}}{d-1} {\cal A}_d (m \ell^d)^2\label{genvol}\label{deltavol}\\
& & {\cal F} = \frac{\pi^{\frac{1}{2}}  (d-1)^{2} \Gamma(d+1)}{G_{(d+1)} 2^{d+6} (d+1) \Gamma\left(d+\frac{3}{2}\right) R_1 {\cal A}_d}\Delta V^{(2)}\label{fidelity}
\eea
where ${\cal A}_d$ is a $d$-dependent constant. Now comparing the expression \eqref{genvol} for $d=4$ with \eqref{changevol} we easily
identify ${\cal A}_4 = \frac{3\pi}{1280}$. Then the fidelity for the AdS$_5$ black hole can be calculated from \eqref{fidelity} for $d=4$ and \eqref{changevol} as,
\be
{\cal F}_{BH} = \frac{ \pi}{525 G_5}R_1^3 m^2 \ell^8
\ee
The corresponding Fisher information metric for the AdS$_5$ black hole w.r.t $\lambda= m \ell^{4}$ therefore takes the form
\be\label{sarkar}
\mathcal{G}_{{\cal F}_{BH},\lambda} = \partial_\lambda^2{\cal F}_{BH} = \frac{2\pi}{525 G_5} R_1^3 
\ee
 
Here, we would like to remark that the definition of fidelity in \cite{Banerjee:2017qti} contains a $d$-dependent constant ${\cal A}_d$ in the denominator (see eq.\eqref{fidelity}). This is deliberately 
chosen such that it precisely cancels the same constant in $\Delta V^{(2)}$ and the result coincides
with that obtained from the relative entropy calculation given in \cite{Blanco:2013joa, Lashkari:2015hha} for AdS black hole. However, we can directly calculate the Fisher information from the second order change in EE using the definition given in \cite{Blanco:2013joa} as, 
\begin{equation}\label{fisherbh}
 G_{{\cal F}_{BH},\lambda} = \frac{d^2}{d \lambda^{2}} (\Delta H - \Delta S_{EE(BH)}) = -\frac{2}{\lambda^{2}} \Delta S_{EE(BH)}^{(2)}.
\end{equation}
In the above, $\lambda\, (= m\ell^4)$ is the perturbation parameter and $\Delta H$ is the change of modular Hamiltonian, which is defined in information theory as $H = -\log\rho$, where $\rho$ is the density matrix. One can show that in case of spherical subregion, $\Delta H$ is only first order in $\lambda$ (all higher order changes are zero) and equal to $\Delta E$. Using this fact and the above identity, we find a relation between Fisher Information and the change in volume in general dimension (given in our recent paper \cite{Bhattacharya:2019zkb}) as,
\begin{equation}\label{fisherbh1}
 G_{{\cal F}_{BH},d,\lambda} =  \frac{ (d-1)^2 \Gamma (d+1) \Gamma \bigl( \frac{d}{2} -1 \bigr)}{8 G_{d+1} \lambda^2 R_{1}  \Gamma \bigl( d + \frac{3}{2} \bigr) \Gamma \bigl( \frac{d+1}{2} \bigr)} \Delta V_{(BH),d}^{(2)}
\end{equation}
For $d=4$, this relation reduces to,
\begin{equation}\label{fisherbh1ads5}
 G_{{\cal F}_{BH},\lambda} =  \frac{128}{105 \pi R_{1} \lambda^{2} G_{5}} \Delta V_{(BH)}^{(2)} = 
 \frac{2\pi}{525 G_5} R_1^3,
\end{equation}
where we have used \eqref{changevol} for $\Delta V^{(2)}$. This precisely matches with the expression we obtained in \eqref{sarkar} according to the definition used in \cite{Banerjee:2017qti}. We will however
find that these two ways of calculating the Fisher information metric yield different results for the
decoupled geometry of `black' non-susy D3 brane to which we turn in the next section.

\section{EE and complexity for (decoupled) `black' non-susy D3 brane in case of spherical subsystem}

The decoupled geometry of `black' non-susy D3 brane is given in eq.\eqref{aads5}. To compare our results with those of the previous section
we will not directly use this geometry, but instead try to recast the solution in a form very similar to the AdS$_5$ black hole geometry. For
this purpose we first make a coordinate transformation $\tilde{\rho}^{4}=\rho^{4} + \rho_0^4$. Then we make another coordinate transformation 
by taking $\tilde{\rho}=\frac{R_{1}^{2}}{z}$. With these transformations the decoupled geometry of non-susy `black' D3 brane \eqref{aads5}
takes the form 
\begin{equation}\label{nonsusymetric}
ds^{2}= \frac{R_{1}^{2}}{z^{2}}\left[-(1-mz^4)^{\frac{1}{4} - \frac{3\d_2}{8}} dt^2 + (1-m z^{4})^{\frac{1}{4}+\frac{\d_{2}}{8}}\sum_{i=1}^{3} (dx_{i})^{2} + \frac{dz^{2}}{(1-m z^{4})}\right]
\end{equation}
where $m = \frac{1}{z_0^4}$ and $z_{0}=\frac{R_{1}^{2}}{\rho_{0}}$.
To compute the entanglement entropy, the complexity and the associated Fisher information metric for the decoupled geometry of `black' non-susy
D3 brane we will use the metric given in \eqref{nonsusymetric} with choice of a spherical subsystem.
We remark that as the Ryu-Takayanagi area or the volume formula uses the Einstein-frame metric, we also use the Einstein frame metric for the decoupled `black' non-susy D3 brane geometry. This, in turn, takes into account that we  have a  non-trivial dilaton in the background.  
The area integral,
after taking the embedding $r=r(z)$, in this case 
takes the form, 
\be\label{area3}
A_{nsD3} = 4 \pi R_{1}^{3}\int dz \frac{r(z)^{2}(1-m z^{4})^{\frac{\d_{2}}{8}-\frac{1}{4}}}{z^{3}}\left[ 1+(1-m z^{4})^{\frac{5}{4}+\frac{\d_{2}}{8}}r^{\prime}(z)^{2}  
\right]^{\frac{1}{2}}
\ee
Again, as before, we are assuming the small subsystem and consider upto the second order change in the metric.
To minimize this area, we use Euler-Lagrange equation of motion once we consider the area as action integral.
The equation of motion is a bit long and so we do not write it explicitly here. We just give its solution. As mentioned earlier, we know that by taking $m=0$, 
we can get back the pure AdS$_5$ case. Thus we again take our solution as a perturbation over pure AdS$_5$ and as before work with the ansatz
\be
r(z) = \sqrt{L^2-z^2} + m r_{1}(z) + m^2 r_{2}(z).
\ee
Now solving the equation of motion with this ansatz, and with proper boundary conditions and regularity conditions 
(similar to the AdS$_5$ black hole case discussed in the previous section), we get 
$r_{1}(z)$ and $r_{2}(z)$ to be of the form
\bea\label{r1r2}
& & r_{1}(z)=\frac{1}{80} \sqrt{L^2-z^2} \left[(10-3 \d_{2}) L^4+(10-3 \d_{2}) L^2 z^2+(\d_{2}+10) z^4\right],\nn
& & r_{2}(z)= \frac{1}{806400}\sqrt{L^2-z^2}\left[(\d_{2} (5207 \d_{2}-18900)+30460) L^8\right.\nn
& & \qquad\qquad +80(\d_{2}(58\d_{2}-189)+302) L^6 z^2 +3 (\d_{2} (683 \d_{2}-2100)+7660) L^4 z^4\nn
& & \qquad\qquad \left. +8 (\d_{2} (263 \d_{2}-1260)+4300) L^2 z^6 +175 (\d_{2}+10) (\d_{2}+26) z^8 \right]
\eea

As before, we can now use this form of $r(z)$ to get the relation between $\ell$ and $L$, but, what we really need is the inverse of that.
This comes out as,
\begin{equation}\label{Ll1}
 L= \ell+ \frac{1}{80} m\ell^5 (3 \d_{2} -10)+m^2\ell^9 \frac{ \left(463 \d_{2}^2 -18900 \d_{2} +32540 \right)}{806400}
\end{equation}
Using the form of $r(z)$ alongwith \eqref{r1r2} in the area integral \eqref{area3} and then
expanding the integral in the second order in $m$ in the way we mentioned before, we perform the integrals upto order $m^{2}$.

After performing the integral (as done before) and replacing $L$ by \eqref{Ll1}, we get the first and second order change of EE with respect 
to $m$ as, 
\bea\label{entropynonsusy1}
    & & \Delta S_{EE(nsD3)}^{(1)} = -\frac{\pi \d_{2}R_{1}^{3}}{20 G_{5} }m \ell^{4}\\
& & \Delta S_{EE(nsD3)}^{(2)} = \frac{\left(\d_{2}^2-10\right)\pi R_{1}^{3}}{3150 G_{5} }m^2 \ell^{8}\label{entropynonsusy2}
\eea
Note that both of these match precisely with the change in EE we obtained for AdS$_5$ black hole in \eqref{entropy1} and \eqref{entropy2} once we put $\d_2=-2$
and provides a consistency check of our result \eqref{entropynonsusy1} and \eqref{entropynonsusy2}. 

Now to compute the complexity we have to find the RT volume from the geometry given in \eqref{nonsusymetric}. 
The volume integral has the form, 
\begin{equation}
    V_{nsD3}=\frac{4\pi R_{1}^{4}}{3}\int_{\epsilon}^{L}\frac{dz}{z^4} r(z)^{3}\left(1-\frac{ \d_{2}}{8}m z^4\right)^{\left(\frac{\d_{2}}{16}-\frac{1}{8}\right)}
\end{equation}
Putting the functional form of $r(z)$ and expanding up to second order in $m$, we get the integrals upto second order.

Evaluating these integrals and taking $\epsilon \rightarrow 0$ limit, we find that the change of complexity upto first order 
in $m$ is zero similar to the case of AdS$_5$ black hole. On the other hand, the change of complexity in the 
second order in $m$ is
\be\label{complexns}
\Delta C_{V(nsD3)}^{(2)} = \left(\frac{4 \pi R_{1}^{3}}{24 \pi G_{5}}\right)\left[\frac{ \pi  \left(60 - 9\d_2^2\right)}
{10240}\right] (m \ell^{4})^{2} = \frac{\Delta V_{(nsD3)}^{(2)}}{8 \pi R_1 G_{5}}
\ee
This can be seen to match with the change in AdS$_5$ black hole complexity given in \eqref{complexbh} once
we take $\d_{2}=-2$. Now using the $\langle T_{tt}\rangle$ calculated in \cite{Bhattacharya:2017gzt}, the change in energy for the non-susy geometry can be obtained as
\be
\langle T_{tt} \rangle = \frac{-3 R_1^3 \d_2}{32 \pi G_5} ,\qquad  \Delta E = \frac{4\pi \ell^3}{3}\langle T_{tt}\rangle = - \frac{\d_2 R_1^3 m \ell^3}{8 G_5}.
\ee

Thus we see that again we can write the change in EE in the  form $\Delta S_{EE(nsD3)}^{(1)} = \frac{\Delta E}{T_{ent(nsD3)}}$, where
$T_{ent(nsD3)} = \frac{5}{2\pi \ell}$. Note that the entanglement temperature remains the same as for the AdS$_5$ black hole. 
Similarly, we can express the change in complexity
\eqref{complexns} as, 
\be
\Delta C_{V(nsD3)}^{(2)} = \frac{5\left(60 - 9\d_2^2\right)G_5}{256 \pi \d_{2}^2 R_1^3}\left(\frac{\Delta E}{T_{ent(nsD3)}}\right)^2
\ee
We also compute the fidelity and Fisher information metric for the non-susy geometry.
Comparing this with the general expression of change of volume \eqref{deltavol},
we identify the $d$-dependent constant ${\cal A}_4$ and fidelity in this case as,
\be
{\cal A}_4 = \frac{ \pi  \left(60 - 9\d_2^2\right)}{10240}, \qquad  {\cal F}_{nsD3} = \frac{ \pi}{525 G_5} R_1^3 m^2 \ell^8
\ee
The corresponding Fisher information metric has the form
\be
\mathcal{G}_{{\cal F}_{nsD3},\lambda} = \partial_\lambda^2 {\cal F}_{nsD3} = \frac{2 \pi}{525 G_5}R_1^3 
\ee  
Interestingly, here we observe that both the fidelity and the Fisher information metric do not depend on the non-susy parameter $\d_2$ and by comparison we see that they have exactly the same value as those of the AdS$_5$ black hole. But this is due to the 
choice of the $d$-dependent constant ${\cal A}_d$ in the denominator of the fidelity used in \cite{Banerjee:2017qti}. 
Next, we consider the direct way of calculating the Fisher information metric as discussed in the previous section. 
Using the definition given in \eqref{fisherbh} and also the relation for $\Delta V^{(2)}_{(nsD3)}$ in \eqref{complexns}
we get,
\begin{eqnarray}\label{im}
G_{{\cal F}_{nsD3},\lambda} &=& -\frac{2}{\lambda^{2}} \Delta S_{EE(nsD3)}^{(2)}\nonumber\\
&=& \frac{512 \left(10-\d_{2}^2\right)}{105 \pi  \left(60-9 \d_{2}^2\right) \pi R_1 \lambda^{2} G_{5} } \Delta V_{(nsD3)}^{(2)}\nonumber\\ 
&=& \frac{ \left(10 - \d_{2}^2 \right)\pi R_{1}^{3}}{1575 G_{5} }.
\end{eqnarray}
Here we find that the Fisher information metric indeed depends on the non-supersymmetric parameter $\d_{2}$ which at $\d_{2} = -2$
gives back the AdS black hole result $G_{{\cal F}_{BH},\lambda}$. 

It is, therefore, clear that the definition of fidelity, used in \cite{Banerjee:2017qti} which contains the $d$-dependent constant ${\cal A}_d$ needed to get the correct AdS black hole result for Fisher information metric, does not produce the correct result for the non-supersymmetric background. This calculation gives the Fisher information metric which is independent of the non-supersymmetric parameter $\d_2$ and has precisely the same value as that of the AdS$_5$ black hole.
However, a direct way of calculating the Fisher information metric given in \cite{Blanco:2013joa}, yields a different result and in this case it depends on the non-supersymmetric parameter $\d_2$ as expected and for $\d_2=-2$, it gives the correct AdS$_5$ black hole result. We observe from \eqref{im} that the decoupled `black' non-susy D3 brane geometry stores more quantum Fisher information than
its AdS$_5$ black hole counterpart. For $\d_2=0$ which corresponds to the zero temperature non-susy solution, in fact, stores the most quantum Fisher information, whereas for $\d_2=-2$, which corresponds to the AdS$_5$ black hole
stores the least.

\section{Conclusion}

To conclude, in this paper we have holographically computed the EE and the complexity of the QFT whose 
gravity dual is given by the decoupled geometry of `black' non-susy D3 brane of type IIB string theory for spherical subsystems.
We start with a brief review of the computation of holographic EE and the complexity of the AdS$_5$ black hole for
spherical subregion done before (complexity was done explicitly for AdS$_3$ and AdS$_4$ in \cite{Alishahiha:2015rta}) 
and then compute the entanglement entropy and subregion complexity for the decoupled `black' non-susy D3 brane geometry 
upto the second order in perturbation paramater using the prescription of Ryu and 
Takayanagi. We have extended our calculation of complexity to compute the fidelity and the Fisher information
metric using the definition given earlier \cite{Banerjee:2017qti} for both the AdS$_5$ black hole
and the decoupled `black' non-susy D3 brane geometry for 
spherical subsystem. Since the decoupled geometry of `black' non-susy D3 brane reduces to the standard AdS$_5$ black hole when
its parameter $\d_2$ takes value $-2$, we have observed that both the EE and the complexity for the former geometry indeed
reduce to those of the AdS$_5$ black hole when we put $\d_2=-2$, giving a consistency check of our results.
We have also checked the entanglement thermodynamics to be consistent for the spherical subsystem and gives the 
same entanglement temperature as the AdS$_5$ black hole. We have further observed that although the fidelity and
the Fisher information metric of the QFT dual to decoupled `black' non-susy D3 brane geometry remain the same as those of
the AdS$_5$ black hole when one uses proposal of \cite{Banerjee:2017qti}, using a more exact relation \cite{Blanco:2013joa} without any arbitrary
constant gives us a different value of Fisher information in case of the non-supersymmetric solution, which is parameter
dependent. Putting the right parameter value gives back the AdS$_5$ black hole result, indicating \eqref{im} is a more general 
relation which includes the AdS$_5$ black hole relation \eqref{fisherbh1} as well.

\bibliographystyle{JHEP}
\bibliography{bhatroy.bib}
\end{document}